\documentclass[%
aip,
pop,
amsmath,amssymb,
reprint,%
]{revtex4-1}

\usepackage{graphicx}
\usepackage{dcolumn}
\usepackage{bm}
\usepackage{color}

\usepackage[utf8]{inputenc}
\usepackage[T1]{fontenc}
\usepackage{mathptmx}
\usepackage{gensymb}

\renewcommand{\vec}[1]{\mathbf{#1}}
\newcommand{\abs}[1]{\left| #1 \right|} 
\begin{document}


\title{Dispersion relation of square lattice waves in a two-dimensional binary complex plasma}

\author{Z.-C. Fu}
 \affiliation{College of Science, Donghua University, 201620 Shanghai, People’s Republic of China}
\author{A. Zampetaki}
\affiliation{Max-Planck-Institut f\"{u}r Extraterrestrische Physik, 85741 Garching, Germany}%
\affiliation{Institut f\"ur Theoretische Physik II: Weiche Materie, Heinrich-Heine-Universit\"at D\"usseldorf, 40225 D\"usseldorf, Germany}%
\author{H. Huang}
\affiliation{College of Science, Donghua University, 201620 Shanghai, People’s Republic of China}%
\author{C.-R. Du}
\email{chengran.du@dhu.edu.cn}
\affiliation{College of Science, Donghua University, 201620 Shanghai, People’s Republic of China}%
\affiliation{Member of Magnetic Confinement Fusion Research Centre, Ministry of Education, 201620 Shanghai, People’s Republic of China}

\date{\today}

\begin{abstract}
Binary complex plasmas consist of microparticles of two different species and can form two-dimensional square lattices under certain conditions. The dispersion relations of the square lattice waves are derived for the longitudinal and transverse in-plane modes, assuming that the out-of-plane mode is suppressed by the strong vertical confinement. The results are compared with the spectra obtained in Langevin dynamics simulations. Furthermore, we investigate the dependence of the dispersion relation on the charge ratio and mass ratio of the two particle species.
\end{abstract}


\maketitle

Complex plasmas consist of a mixture of weakly ionized gases and  microparticles. The latter acquire a charge due to the flow of the surrounding ions and electrons which are negative, owing to the higher thermal velocity of electrons \cite{fortov:2005}. Considering the plasma screening effect, the interaction between the microparticles can be described via the Yukawa potential \cite{shukla:book}. In the laboratory, the charged particles are usually suspended in the sheath above the lower electrode of a radio-frequency (rf) discharge, where the gravity force is balanced by the electric force. In strongly coupled complex plasmas, monodisperse microparticles can be vertically confined to a single layer and form a hexagonal lattice, known as plasma crystal \cite{i:1996,thomas:1996}. Due to the stretched time scales and low damping, two-dimensional (2D) complex plasma crystals provide an unique opportunity to study generic processes in solids and liquids at the kinetic level \cite{Morfill:2009}. With external manipulations by electric fields or laser beams, various phenomena such as melting \cite{feng:2010} and recrystallization \cite{knapek:2007,nosenko:2013}, microstructure under shear \cite{nosenko:2012}, Mach cone excitations \cite{melzer:2000}, and entropy production \cite{wong:2018,wieben:2019} have been investigated  both experimentally and theoretically. 

One of the most defining properties of plasma crystals is the dispersion relation of the microparticles' collective oscillations, in the  form of lattice waves. This has been derived analytically and measured directly using video microscopy in the case of monodisperse complex plasmas \cite{wang:2001,nunomura:2002a,nunomura:2002b}. Remarkably, due to the strong ion flow in the sheath, the interactions between microparticles are altered by the so-called wake effect, resulting in the coupling of the horizontal and vertical modes \cite{zhdanov:2009,liu:2010,couedel:2011,roecker:2012,meyer:2017}. This eventually triggers a mode-coupling instability (MCI) and causes the crystal to melt \cite{williams:2012,ivlev:2015a}.

A binary complex plasma consists of microparticles of two different species. With an appropriate selection of their mass and size, these particles can form, in the laboratory, a bilayer \cite{hartman:2009} or a quasi-two-dimensional (q2D) crystalline suspension \cite{ivlev:2015b,du:2019} \footnote{The size of the vertical gap between two particle species is similar to the interparticle distance in the bilayer suspension. In contrast, this gap size is far less than the interparticle distance in the q2D binary complex plasma.}. The phonon spectra for these structures have been measured experimentally and studied by a quasi-localized charge approximation approach and molecular dynamics simulations \cite{kalman:2013,huang:2019}. A mode coupling between the horizontal modes in the two layers, mediated by the plasma wakes, has been proposed \cite{ivlev:2017}. 

\begin{figure}[!hb]
	\includegraphics[scale=0.42]{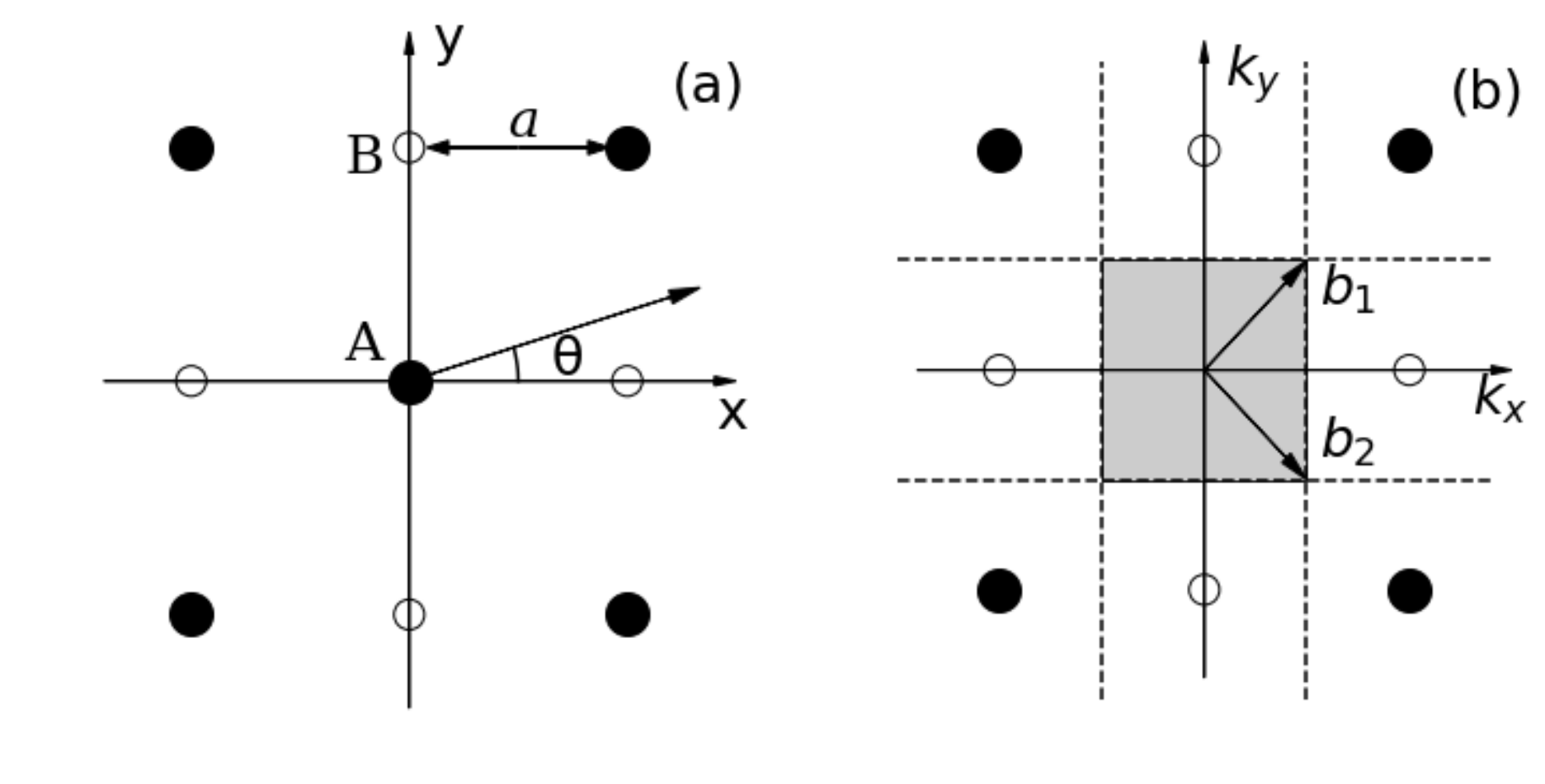}
	\caption{(a) The sketch of a square lattice with particle spacing $a$ in a 2D binary complex plasma consisting of the particle species $A$ and $B$. (b) The reciprocal lattice in the $\textbf{k}$-space, where the basis vectors are $\textbf{b}_{1,2}=2{\pi}a^{-1}(1/2,\pm1/2)$. Due to the square symmetry, we consider the wave vectors $\textbf{k}$ at $0\degree$ and $45\degree$ in the first Brillouin zone (the shaded area surrounded by a dashed line).}
	\label{Fig1}
\end{figure}

Meanwhile, taking advantage of the plasma etching effect, the two particle species can be suspended at the same height for a certain amount of time \cite{wieben:2017}. Under these conditions, binary complex plasmas have been found  to form square lattices with a quadruple symmetry  \cite{ivlev:book,huang:2019} as the one presented in Fig.~\ref{Fig1}. A strong vertical confinement can efficiently suppress the vertical motions and thus the expected MCI \cite{williams:2012,ivlev:2015a}. A natural question that then arises is  how the horizontal wave modes are modified in  such 2D square lattices compared to the  well-studied case of  hexagonal lattices in monodisperse 2D complex plasma crystals.

In this article, we derive the dynamical matrix of a 2D binary complex plasma crystal in order to study the dispersion relation of the formed square lattice and to investigate its dependence on the involved particle charges and masses. In our approach we assume a very strong confinement, resulting in the particles being confined on a 2D plane, and allowing us to neglect the effect of the plasma wakes. 

The interaction between the charged microparticles  is assumed to be of a Yukawa form. Thus, each particle $i$ of species $p=A$ or $B$ and a charge $Q_p$, is a source of a Yukawa potential 
\begin{equation}
	\varphi_{p,i}\left(\vec{r}\right)=\frac{Q_p}{\abs{\vec{r}-\vec{r}_{p,i}}} \exp\left(-\abs{\vec{r}-\vec{r}_{p,i}}/\lambda_D\right) \label{pot1}
\end{equation}
with a screening length $\lambda_D$ at its  2D position $\vec{r}_{p,i}=x_{p,i}\widehat{\vec{x}}+y_{p,i}\widehat{\vec{y}}$. Consequently the equations of motion for the $i$-th particle of species $A$ and the $i$-th particle of species $B$, in the 2D plane of confinement, read
\begin{equation}
\begin{split}
	\frac{d^2 \vec{r}_{A,i}}{dt^2} + \nu \frac{d\vec{r}_{A,i}}{dt} =  -\frac{Q_A}{M_A}  \nabla_{A,i}\left(\sum\limits_{j \neq i} \varphi_{A,j}\left(\vec{r}_{A,i}\right)+  \sum\limits_{j} \varphi_{B,j}\left(\vec{r}_{A,i}\right)\right), \\
	\frac{d^2 \vec{r}_{B,i}}{dt^2} + \nu \frac{d\vec{r}_{B,i}}{dt} =  -\frac{Q_B}{M_B}  \nabla_{B,i}\left(\sum\limits_{j} \varphi_{A,j}\left(\vec{r}_{B,i}\right)+  \sum\limits_{j \neq i} \varphi_{B,j}\left(\vec{r}_{B,i}\right)\right), 
\end{split} \label{eom1}
\end{equation}
where $\nu$ is the frictional drag coefficient and $M_{A,B}$ and $Q_{A,B}$ are the mass and the charge of the particle species $A$ and $B$, respectively. Note also that  the symbols $\nabla_{A,i}$, $\nabla_{B,i}$ denote the gradients with respect to the vectors  $\vec{r}_{A,i}$ and $\vec{r}_{B,i}$, accordingly.

Our scope here is to investigate the vibrational properties of the 2D binary complex plasma crystal around its square lattice equilibrium, shown in Fig. \ref{Fig1}.  In this crystalline configuration, the microparticles of the species $p=A$ or $B$ are located at the positions
$\vec{R}_p=X_p\widehat{\vec{x}}+Y_p\widehat{\vec{y}}$  with $X_A=(m+n)a$,
 $Y_A=(m-n)a$, $X_B=1-X_A$ and $Y_B=Y_A$ where $m,n$ are arbitrary integers and $a$ 
denotes the square  lattice constant (Fig. \ref{Fig1}). Linearizing around this equilibrium and introducing the plane wave ansatz 
\begin{equation}
	\vec{d}_{p}=\vec{d}_{p}^{(0)} \exp\left[-i\omega t+i \left(k_x X_p+k_y Y_p\right) \right]
\end{equation}	
for the displacement $\vec{d}_{p}$ of the particles of species $p$ from their equilibrium positions $\vec{R}_p$,
we arrive, in view of the square lattice symmetry, at the $4\times 4$ dynamical matrix
\begin{widetext}
\begin{equation}
	{\vec{D}}={\frac{Q_A^2}{M_A}}
	\left(
	\begin{array}{cccc}
	F_{A}^s + {\frac{1}{\Lambda_Q}}F_{B}^o & G_{A}^s + {\frac{1}{\Lambda_Q}}G_{B}^o 
	& -{\frac{1}{\Lambda_Q}}F_{B}^l  & -{\frac{1}{\Lambda_Q}}G_{B}^l \\
	
	G_{A}^s + {\frac{1}{\Lambda_Q}}G_{B}^o   & \overline{F}_{A}^s + {\frac{1}{\Lambda_Q}}\overline{F}_{B}^o 
	& -{\frac{1}{\Lambda_Q}}G_{B}^l & -{\frac{1}{\Lambda_Q}}\overline{F}_{B}^l \\
	
	-{\frac{\Lambda_M}{\Lambda_Q}}F_{B}^l & -{\frac{\Lambda_M}{\Lambda_Q}}G_{B}^l & {\frac{\Lambda_M}{\Lambda_Q^2}}F_{A}^s + {\frac{\Lambda_M}{\Lambda_Q}}F_{B}^o & {\frac{\Lambda_M}{\Lambda_Q^2}}G_{A}^s + {\frac{\Lambda_M}{\Lambda_Q}}G_{B}^o \\
	
	-{\frac{\Lambda_M}{\Lambda_Q}}G_{B}^l & -{\frac{\Lambda_M}{\Lambda_Q}}\overline{F}_{B}^l & {\frac{\Lambda_M}{\Lambda_Q^2}}G_{A}^s + {\frac{\Lambda_M}{\Lambda_Q}}G_{B}^o & {\frac{\Lambda_M}{\Lambda_Q^2}}\overline{F}_{A}^s + {\frac{\Lambda_M}{\Lambda_Q}}\overline{F}_{B}^o
	\end{array}
	\right).
\end{equation}
\end{widetext}
In this expression  we have denoted the mass ratio as ${\Lambda_M}={M_A}/{M_B}$ and the charge ratio as ${\Lambda_Q}={Q_A}/{Q_B}$. The elements $F_p^o$, $F_p^l$, and $F_p^s$ with $p=A$ or $B$  are given by the following sums of the effective spring constant $F(X,Y)$ over the lattice positions $(X_p,Y_p)$, excluding the central position:  
\begin{equation}
\begin{split}
	F_p^o &= \sum\limits_{X_p,Y_p } F(X_p,Y_p), \\
	F_p^l &= \sum\limits_{X_p,Y_p} F(X_p,Y_p) \cos (k_x X_p + k_y Y_p), \\
	F_p^s &= F_p^o-F_p^l.
\end{split}
\end{equation}
Such summations also apply to the rest of the matrix elements $\overline{F}_p^{o,l,s}$ and $G_p^{o,l,s}$. The corresponding effective spring constants read
\begin{equation}
\begin{split}
	F(X,Y)            =& R^{-5}e^{-R/\lambda_D}[X^2 (3+3R/\lambda_D +R^2/\lambda_D^2) \\ 
                       & -R^2(1+R/\lambda_D)] , \\
    \overline{F}(X,Y) =& F(Y,X), \\
    G(X,Y)            =& (XY/R^5)e^{-R/\lambda_D}(3+3R/\lambda_D +R^2/\lambda_D^2) \\
\end{split}
\end{equation}
with $R=\sqrt{X^2+Y^2}$.

The eigenvalues $\Omega_j^2$ of the dynamical matrix $\vec{D}$ are connected with the eigenfrequencies $\omega_j$ of the crystal  through the relation $\Omega_j^2=\omega_j \left(\omega_j+i \nu\right)$. For simplicity, we  approximate in the following theoretical results the $\omega_j$ with the respective values of $\Omega_j$, under  the assumption that the damping in our system is very small $(\nu \ll \omega_j)$. Since $\vec{D}$ is a $4 \times 4$ matrix its eigenvalues  yield 4 branches $\Omega_j(k_x,k_y)$, two of which are transverse and two longitudinal.

\begin{figure}[!hb]
	\includegraphics[scale=0.4]{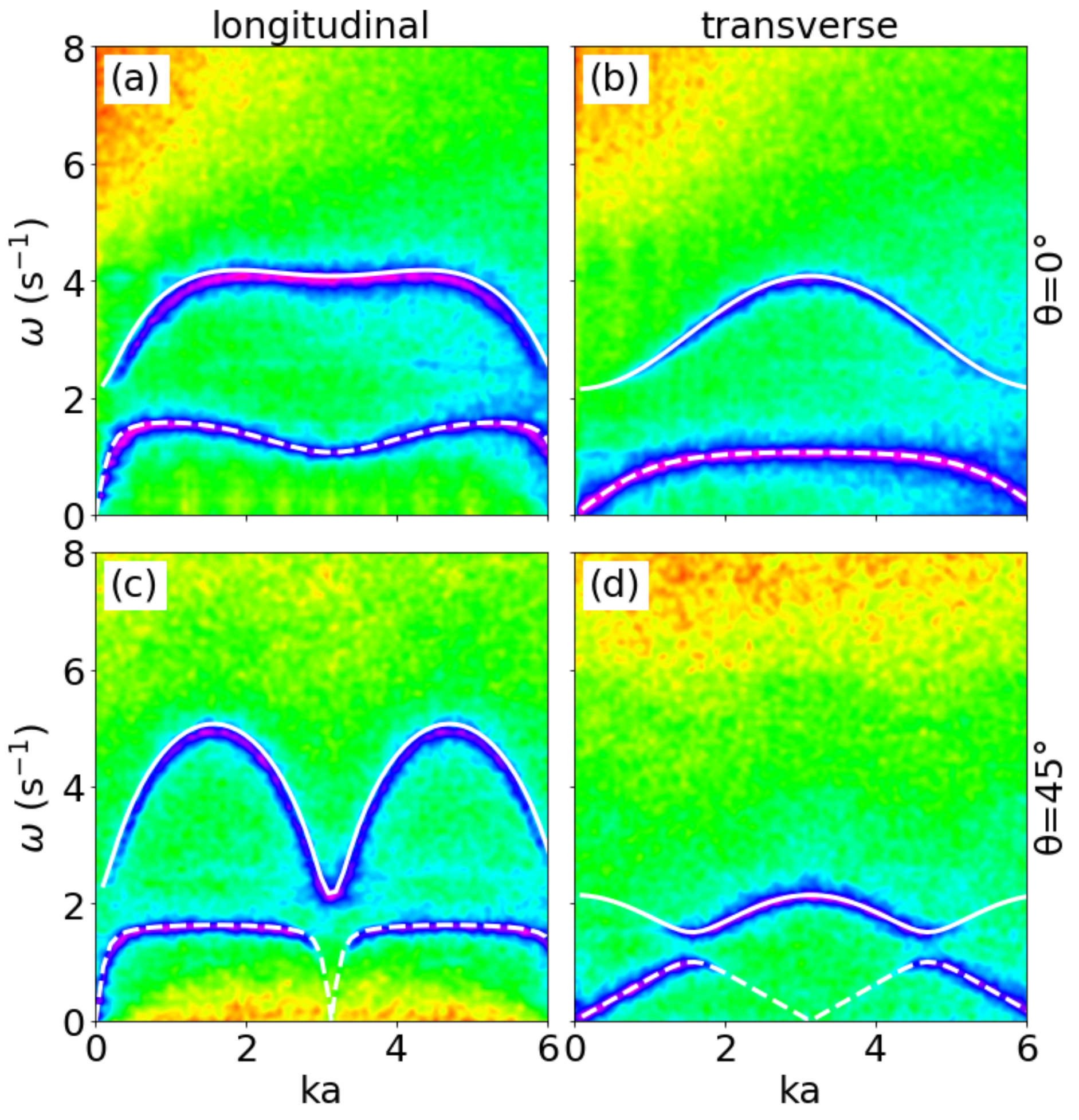}
	\caption{The phonon spectra obtained from the Langevin dynamics simulation (color map) and the theoretically calculated dispersion relations, i.e. the eigenvalues $\Omega_j$ of the dynamical matrix $\vec{D}$ (white curves). Upper panels show (a) the longitudinal and (b) the transverse modes at the wave vector angle $\theta=0 \degree$. Lower panels show (c) the longitudinal and (d) the transverse modes at $\theta=45 \degree$.}
	\label{Fig2}
\end{figure}

An example of these spectra for $\Lambda_M=1$ and $\Lambda_Q=8$,  at the two characteristic wave vector angles for the square lattice, $\theta =0\degree$ and $\theta =45\degree$, is shown in Fig.~\ref{Fig2} (a),(b) and Fig.~\ref{Fig2} (c),(d), respectively. Each of the longitudinal  [Fig.~\ref{Fig2} (a),(c)] and the transverse modes [Fig.~\ref{Fig2} (b),(d)], possesses two branches, due to the two-particle unit cell of the binary square lattice. From these the lower one corresponds to the acoustic branch, with the  two particles of the unit cell oscillating in phase, and the higher one to the optical branch, with the two particles oscillating out of phase.

In order to corroborate our theoretical calculation, we perform a 2D Langevin dynamics simulation with periodic boundary conditions using LAMMPS in the NVT ensemble \cite{plimpton:1995}. In our simulation, the two species of particles are modelled as negative point-like charges and are arranged in a square lattice in a single layer, as illustrated in Fig.~\ref{Fig1}. Note that we do not apply any external excitation to trigger the formation of lattice waves. Instead, we measure the phonon spectra from the collective thermal motions of the particles by applying  a Langevin thermal bath of temperature $T=500$~K with a damping rate $\nu=0.1$~s$^{-1}$. We also set the interparticle distance to $a=0.4$~mm and the screening length to $\lambda_D=0.8$~mm.  The total number of particles used is $N=8100$. The masses of the two particles are assumed to be equal, with $M_A = M_B = 6 \times 10^{ - 13}$~kg, whereas their charges  are set to $Q_A = 8000$~$e$ and $Q_B = 1000$~$e$, respectively. Further details of the simulation can be found in the references \cite{lin:2018,huang:2019}. 

The wave spectra ${\textbf{V}_{{\textbf{k}},\omega}}$ of the simulated square lattices in the binary complex plasma are computed using the 2D Fourier transform
\begin{equation}
   \textbf{V}_{\textbf{k},\omega}=2/ST\int_0^S\int_0^T \textbf{v}(\textbf{r},t)\exp[-i(\textbf{k} \cdot \textbf{r} + \omega t)]dsdt,
\end{equation}
where $S$ and $T$ are the linear size of the area in the simulation and the period over which particle motion is summed, respectively \cite{huang:2019}. As shown in Fig.~\ref{Fig2}, both the acoustic and the optical branches of the wave spectra can be clearly identified in our simulation results. They are also in a very good agreement with our theoretical results using the dynamical matrix approach, indicating that  the  finite temperature and the damping have a marginal influence on the vibrational properties of the system, as long as the square lattice structure  persists.

\begin{figure}[!t]
	\includegraphics[scale=0.4]{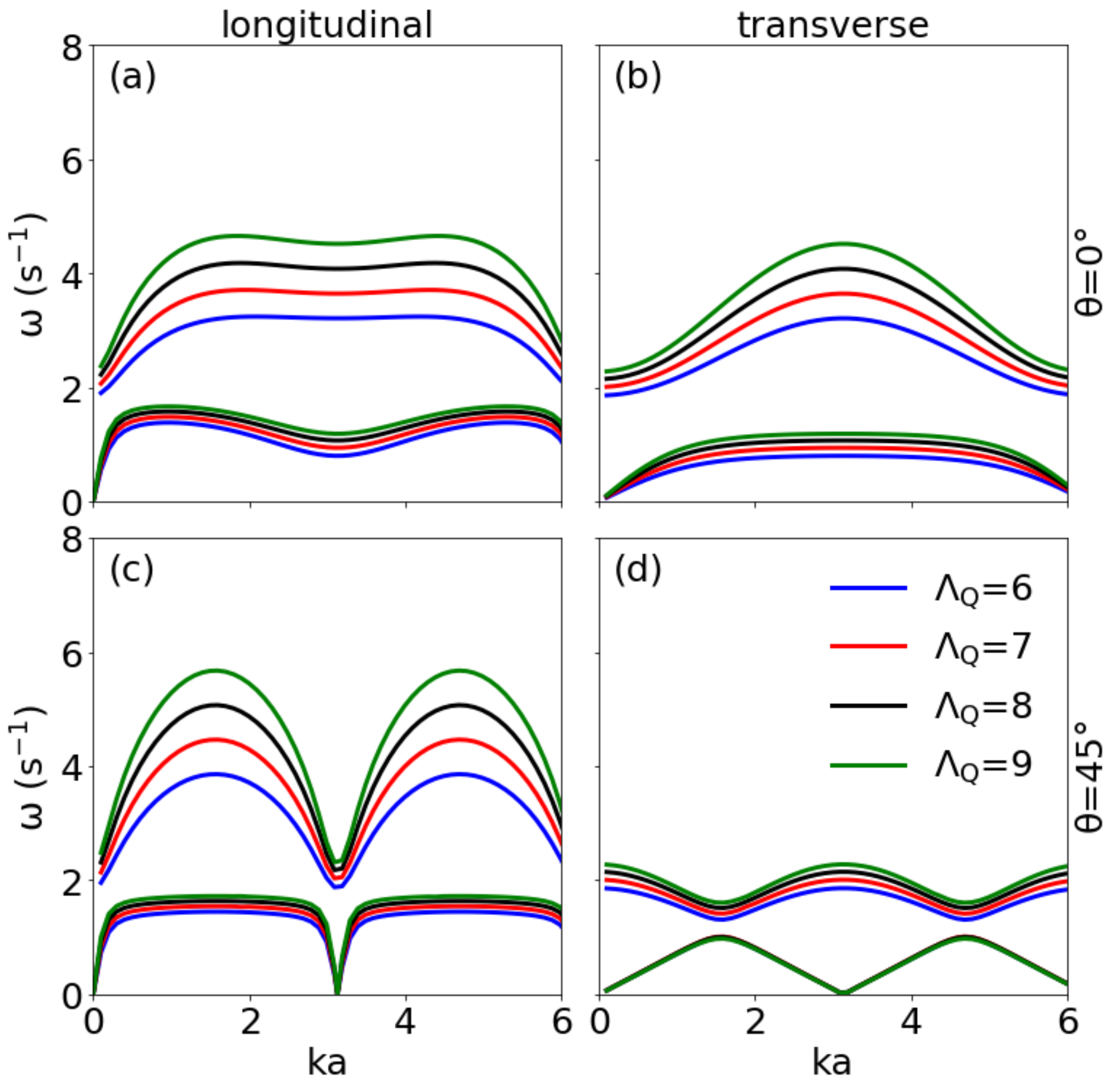}
    \caption{The dependence of the dispersion relation of the binary square crystal on the charge ratio $\Lambda_Q=6$, $7$, $8$ and $9$. Upper panels show (a) the longitudinal and (b) the transverse modes at $\theta=0 \degree$. Lower panels show (c) the longitudinal and (d) the transverse modes at $\theta=45 \degree$. The mass ratio used is $\Lambda_M=1$.}
    \label{Fig3}
\end{figure}
    
\begin{figure}[!t]
	\includegraphics[scale=0.4]{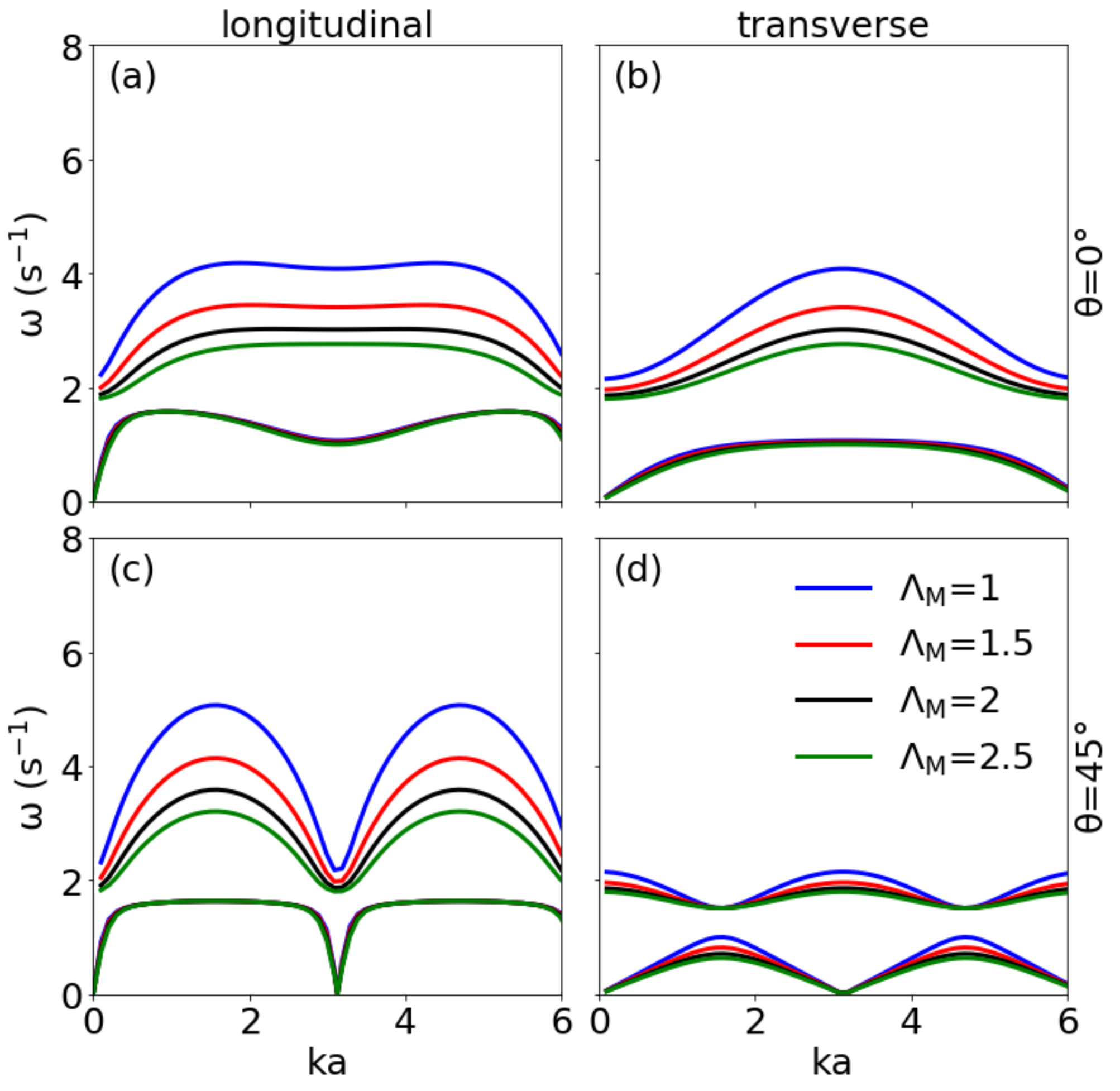}
	\caption{The dependence of the dispersion relation of the binary square crystal on the mass ratios $\Lambda_M=1$, $1.5$, $2$ and $2.5$. Upper panels show (a) the longitudinal and (b) the transverse modes at $\theta=0 \degree$. Lower panels show (c) the longitudinal and (d) the transverse modes at $\theta=45 \degree$, respectively. The charge ratio used is $\Lambda_Q=8$.}
	\label{Fig4}
\end{figure}

In order to study the dependence of the dispersion relation on the disparity of two particle species, we fix the mass and the charge of the particle species $B$ and vary  in our theoretical calculations the mass ratio $\Lambda_M$ and the charge ratio $\Lambda_Q$. The dispersion relation for both the longitudinal and the transverse modes are shown in Fig.~\ref{Fig3} for $\Lambda_M=1$ and different values of $\Lambda_Q$ and in Fig.~\ref{Fig4}  for $\Lambda_Q=8$ and different values of $\Lambda_M$.
	
As we observe in Fig.~\ref{Fig3}, the eigenfrequencies  of the longitudinal and transverse  optical branches increase significantly as the charge ratio $\Lambda_Q$  increases, at both $\theta=0 \degree$ and $\theta=45 \degree$. In contrast, the eigenfrequencies of the acoustic branches increase only moderately. Particularly, the frequencies of the transverse acoustic branch  at $\theta=45 \degree$ [Fig.~\ref{Fig3} (d)]  seem to be hardly affected by the variation of $\Lambda_Q$.

Regarding the dependence of the dispersion relation of the binary complex plasma crystal on the mass ratio $\Lambda_M$ (Fig.~\ref{Fig4}), the frequencies  of all the optical branches decrease as $\Lambda_M$ increases. However, for the acoustic branch, the dispersion relation does not show any dependence on the mass ratio, except for the transverse acoustic branch at $\theta=45 \degree$. For this branch, as depicted in Fig.~\ref{Fig4} (d), the frequencies decrease with  $\Lambda_M$ at the same extent as they do for the optical branch.

To summarize, we have studied the dispersion relations of square lattice waves in a 2D binary complex plasma using the dynamical matrix approach. Both acoustic and optical branches are observed for the longitudinal and transverse modes of the system, as expected for a binary system. The results are found to be in a good agreement with Langevin dynamic simulations, allowing  us to verify our calculations. Furthermore, we have investigated the dependence of the dispersion relation on the charge ratio and the mass ratio of the two particle species. The results provide a comprehensive understanding of the phonon spectra in square lattices of 2D binary complex plasmas.

For the study of the square lattice observed in  quasi-2D complex plasmas \cite{huang:2019,du:2019},  the effect of the ion wakes in the plasma sheath must be explicitly considered, since it alters the particles' interactions, rendering them non-reciprocal \cite{ivlev:2015b}. In the simplest approach the wakes can  be modeled as  point-like positive charges located directly beneath the microparticles \cite{ivlel:2000} and the dispersion relation of the lattice waves can be calculated through the dynamical matrix, following a similar procedure as the one employed in this article. Our results can therefore serve as a benchmark for the more involved calculations for quasi-2D complex plasmas with non-reciprocal interactions, which will be addressed in future works. 

\begin{acknowledgments}
This work is supported by the National Natural Science Foundation of China (NSFC), Grant No. 11975073. The authors would like to thank A. Ivlev for the valuable discussions. 
\end{acknowledgments}

The data that support the findings of this study are available from the corresponding author upon reasonable request.


%

\end{document}